\documentclass[sigconf]{acmart} 

\AtBeginDocument{%
  \providecommand\BibTeX{{%
    \normalfont B\kern-0.5em{\scshape i\kern-0.25em b}\kern-0.8em\TeX}}}

\setcopyright{none}
\copyrightyear{2025}
\acmYear{2025}
\acmDOI{10.1145/XXXXXXX.XXXXXXX}

\acmConference[SBES '25]{Brazilian Symposium on Software Engineering}{September 22--26, 2025}{Recife, PE}

\setcopyright{none} 
\renewcommand\footnotetextcopyrightpermission[1]{} 

\usepackage{listings}
\usepackage{tcolorbox}
\usepackage{geometry}
\usepackage{amsmath}
\usepackage{xcolor}
\usepackage[utf8]{inputenc}
\usepackage{newunicodechar}
\usepackage{stfloats}
\usepackage{minted}
\usepackage{clipboard}
\usepackage{ifthen}
\usepackage{xparse}
\usepackage{xstring}
\usepackage{etoolbox}
\usepackage{pgffor}
\usepackage{tablefootnote}
\usepackage{threeparttable}
\usepackage[justification=centering]{caption}
\tcbuselibrary{listings, skins}
\newcommand{\red}[1]{\textcolor{red}{#1}}

\begin{document}


\newcommand{\llmsEnumString}{gpt, llama, claude, gemini, deepseek, mistral, qwen, sabia}
\edef\llmsEnum{\llmsEnumString}

\newcommand{\llmsIndexEnumString}{1, 2, 3, 4, 5, 6, 7, 8}
\edef\llmsIndexEnum{\llmsIndexEnumString}

\newcommand{\llmsScoreEnumString}{s1, s2, s3, s4, s5, s6, s7, s8}
\edef\llmsScoreEnum{\llmsScoreEnumString}

\newcommand{\llmsSuccessRateEnumString}{p1, p2, p3, p4, p5, p6, p7, p8}
\edef\llmsSuccessRateEnum{\llmsSuccessRateEnumString}

\newcommand{\llmsCoverageEnumString}{c1, c2, c3, c4, c5, c6, c7, c8}
\edef\llmsCoverageEnum{\llmsCoverageEnumString}

\newcommand{\llmsMutationEnumString}{m1, m2, m3, m4, m5, m6, m7, m8}
\edef\llmsMutationEnum{\llmsMutationEnumString}

\NewDocumentCommand{\llm}{m o}{%
  \IfEqCase{#1}{%
    {claude}{%
      \IfNoValueF{#2}{%
        \IfSubStr{#2}{m}{Claude}{}%
        \IfSubStr{#2}{v}{\IfStrEq{#2}{v}{}{~}3.5 Sonnet}{}%
        \IfSubStr{#2}{e}{\IfStrEq{#2}{e}{Anthropic}{~(Anthropic)}}{}%
        \IfStrEq{#2}{p}{USA}{}%
        \IfStrEq{#2}{l}{Private}{}%
        \IfStrEq{#2}{S}{70,9}{}
        \IfStrEq{#2}{P}{100}{}
        \IfStrEq{#2}{C}{71,7}{}
        \IfStrEq{#2}{M}{40,8}{}
        \IfStrEq{#2}{Q}{38,3}{}
        \IfStrEq{#2}{TT}{230}{}
        \IfStrEq{#2}{FT}{0}{}
        \IfStrEq{#2}{TC}{0,47}{}
        \IfStrEq{#2}{FTPL}{0}{}
        \IfStrEq{#2}{FTMI}{0}{}
        \IfStrEq{#2}{FTAU}{0}{}
        \IfStrEq{#2}{FTPR}{0}{}
        \IfStrEq{#2}{FTRC}{0}{}
        \IfStrEq{#2}{FTJD}{0}{}
        \IfStrEq{#2}{PFT}{0,0}{}
        \IfStrEq{#2}{DS}{--}{}
        \IfStrEq{#2}{DP}{--}{}
        \IfStrEq{#2}{DC}{--}{}
        \IfStrEq{#2}{DM}{--}{}
      }%
    }%
    {gpt}{%
      \IfNoValueF{#2}{%
        \IfSubStr{#2}{m}{GPT}{}%
        \IfSubStr{#2}{v}{\IfStrEq{#2}{v}{}{~}4o}{}%
        \IfSubStr{#2}{e}{\IfStrEq{#2}{e}{OpenAI}{~(OpenAI)}}{}%
        \IfStrEq{#2}{p}{USA}{}%
        \IfStrEq{#2}{l}{Private}{}%
        \IfStrEq{#2}{S}{63,4}{}
        \IfStrEq{#2}{P}{98,0}{}
        \IfStrEq{#2}{C}{59,3}{}
        \IfStrEq{#2}{M}{32,9}{}
        \IfStrEq{#2}{Q}{21,8}{}
        \IfStrEq{#2}{TT}{131}{}
        \IfStrEq{#2}{FT}{2}{}
        \IfStrEq{#2}{TC}{0,25}{}
        \IfStrEq{#2}{FTPL}{0}{}
        \IfStrEq{#2}{FTMI}{0}{}
        \IfStrEq{#2}{FTAU}{0}{}
        \IfStrEq{#2}{FTPR}{\red{2}}{}
        \IfStrEq{#2}{FTRC}{0}{}
        \IfStrEq{#2}{FTJD}{0}{}
        \IfStrEq{#2}{PFT}{1,5}{}
        \IfStrEq{#2}{DS}{-7,5}{}
        \IfStrEq{#2}{DP}{-2,0}{}
        \IfStrEq{#2}{DC}{-12,4}{}
        \IfStrEq{#2}{DM}{-7,9}{}
      }%
    }%
    {mistral}{%
      \IfNoValueF{#2}{%
        \IfSubStr{#2}{m}{Mistral}{}%
        \IfSubStr{#2}{v}{\IfStrEq{#2}{v}{}{~}Large}{}%
        \IfSubStr{#2}{e}{\IfStrEq{#2}{e}{Mistral}{~(Mistral)}}{}%
        \IfStrEq{#2}{p}{France}{}%
        \IfStrEq{#2}{l}{Open-source}{}%
        \IfStrEq{#2}{S}{62,2}{}
        \IfStrEq{#2}{P}{98,0}{}
        \IfStrEq{#2}{C}{63,0}{}
        \IfStrEq{#2}{M}{25,5}{}
        \IfStrEq{#2}{Q}{43,3}{}
        \IfStrEq{#2}{TT}{260}{}
        \IfStrEq{#2}{FT}{5}{}
        \IfStrEq{#2}{TC}{0,31}{}
        \IfStrEq{#2}{FTPL}{\red{1}}{}
        \IfStrEq{#2}{FTMI}{0}{}
        \IfStrEq{#2}{FTAU}{\red{3}}{}
        \IfStrEq{#2}{FTPR}{0}{}
        \IfStrEq{#2}{FTRC}{\red{1}}{}
        \IfStrEq{#2}{FTJD}{0}{}
        \IfStrEq{#2}{PFT}{1,9}{}
        \IfStrEq{#2}{DS}{-8,7}{}
        \IfStrEq{#2}{DP}{-2,0}{}
        \IfStrEq{#2}{DC}{-8,7}{}
        \IfStrEq{#2}{DM}{-15,3}{}
      }%
    }%
    {llama}{%
      \IfNoValueF{#2}{%
        \IfSubStr{#2}{m}{LLaMA}{}%
        \IfSubStr{#2}{v}{\IfStrEq{#2}{v}{}{~}3.2 90b}{}%
        \IfSubStr{#2}{e}{\IfStrEq{#2}{e}{Meta}{~(Meta)}}{}%
        \IfStrEq{#2}{p}{USA}{}%
        \IfStrEq{#2}{l}{Open-source}{}%
        \IfStrEq{#2}{S}{63,9}{}
        \IfStrEq{#2}{P}{97,5}{}
        \IfStrEq{#2}{C}{66,1}{}
        \IfStrEq{#2}{M}{28,0}{}
        \IfStrEq{#2}{Q}{36,2}{}
        \IfStrEq{#2}{TT}{217}{}
        \IfStrEq{#2}{FT}{7}{}
        \IfStrEq{#2}{TC}{0,02}{}
        \IfStrEq{#2}{FTPL}{\red{6}}{}
        \IfStrEq{#2}{FTMI}{0}{}
        \IfStrEq{#2}{FTAU}{0}{}
        \IfStrEq{#2}{FTPR}{0}{}
        \IfStrEq{#2}{FTRC}{\red{1}}{}
        \IfStrEq{#2}{FTJD}{0}{}
        \IfStrEq{#2}{PFT}{3,2}{}
        \IfStrEq{#2}{DS}{-7,0}{}
        \IfStrEq{#2}{DP}{-2,5}{}
        \IfStrEq{#2}{DC}{-5,6}{}
        \IfStrEq{#2}{DM}{-12,8}{}
      }%
    }%
    {deepseek}{%
      \IfNoValueF{#2}{%
        \IfSubStr{#2}{m}{Deepseek}{}%
        \IfSubStr{#2}{v}{\IfStrEq{#2}{v}{}{~}R1}{}%
        \IfSubStr{#2}{e}{\IfStrEq{#2}{e}{Deepseek}{~(Deepseek)}}{}%
        \IfStrEq{#2}{p}{China}{}%
        \IfStrEq{#2}{l}{Open-source}{}%
        \IfStrEq{#2}{S}{67,1}{}
        \IfStrEq{#2}{P}{97,0}{}
        \IfStrEq{#2}{C}{67,5}{}
        \IfStrEq{#2}{M}{36,9}{}
        \IfStrEq{#2}{Q}{33,7}{}
        \IfStrEq{#2}{TT}{202}{}
        \IfStrEq{#2}{FT}{4}{}
        \IfStrEq{#2}{TC}{0,78}{}
        \IfStrEq{#2}{FTPL}{\red{1}}{}
        \IfStrEq{#2}{FTMI}{\red{2}}{}
        \IfStrEq{#2}{FTAU}{\red{1}}{}
        \IfStrEq{#2}{FTPR}{0}{}
        \IfStrEq{#2}{FTRC}{0}{}
        \IfStrEq{#2}{FTJD}{0}{}
        \IfStrEq{#2}{PFT}{2,0}{}
        \IfStrEq{#2}{DS}{-3,8}{}
        \IfStrEq{#2}{DP}{-3,0}{}
        \IfStrEq{#2}{DC}{-4,2}{}
        \IfStrEq{#2}{DM}{-3,9}{}
      }%
    }%
    {sabia}{%
      \IfNoValueF{#2}{%
        \IfSubStr{#2}{m}{Sabiá}{}%
        \IfSubStr{#2}{v}{\IfStrEq{#2}{v}{}{~}3}{}%
        \IfSubStr{#2}{e}{\IfStrEq{#2}{e}{Maritaca}{~(Maritaca)}}{}%
        \IfStrEq{#2}{p}{Brazil}{}%
        \IfStrEq{#2}{l}{Private}{}%
        \IfStrEq{#2}{S}{65,5}{}
        \IfStrEq{#2}{P}{97,5}{}
        \IfStrEq{#2}{C}{64,3}{}
        \IfStrEq{#2}{M}{34,7}{}
        \IfStrEq{#2}{Q}{21,7}{}
        \IfStrEq{#2}{TT}{130}{}
        \IfStrEq{#2}{FT}{4}{}
        \IfStrEq{#2}{TC}{0,08}{}
        \IfStrEq{#2}{FTPL}{0}{}
        \IfStrEq{#2}{FTMI}{0}{}
        \IfStrEq{#2}{FTAU}{0}{}
        \IfStrEq{#2}{FTPR}{\red{2}}{}
        \IfStrEq{#2}{FTRC}{0}{}
        \IfStrEq{#2}{FTJD}{\red{2}}{}
        \IfStrEq{#2}{PFT}{3,1}{}
        \IfStrEq{#2}{DS}{-5,4}{}
        \IfStrEq{#2}{DP}{-2,5}{}
        \IfStrEq{#2}{DC}{-7,4}{}
        \IfStrEq{#2}{DM}{-6,1}{}
      }%
    }%
    {qwen}{%
      \IfNoValueF{#2}{%
        \IfSubStr{#2}{m}{Qwen}{}%
        \IfSubStr{#2}{v}{\IfStrEq{#2}{v}{}{~}2.5 32b}{}%
        \IfSubStr{#2}{e}{\IfStrEq{#2}{e}{Alibaba}{~(Alibaba)}}{}%
        \IfStrEq{#2}{p}{China}{}%
        \IfStrEq{#2}{l}{Open-source}{}%
        \IfStrEq{#2}{S}{65,8}{}
        \IfStrEq{#2}{P}{95,5}{}
        \IfStrEq{#2}{C}{68,7}{}
        \IfStrEq{#2}{M}{33,3}{}
        \IfStrEq{#2}{Q}{34,7}{}
        \IfStrEq{#2}{TT}{208}{}
        \IfStrEq{#2}{FT}{8}{}
        \IfStrEq{#2}{TC}{0,09}{}
        \IfStrEq{#2}{FTPL}{\red{4}}{}
        \IfStrEq{#2}{FTMI}{\red{2}}{}
        \IfStrEq{#2}{FTAU}{\red{1}}{}
        \IfStrEq{#2}{FTPR}{0}{}
        \IfStrEq{#2}{FTRC}{0}{}
        \IfStrEq{#2}{FTJD}{\red{1}}{}
        \IfStrEq{#2}{PFT}{3,8}{}
        \IfStrEq{#2}{DS}{-5,1}{}
        \IfStrEq{#2}{DP}{-4,5}{}
        \IfStrEq{#2}{DC}{-3,0}{}
        \IfStrEq{#2}{DM}{-7,5}{}
      }%
    }%
    {gemini}{%
      \IfNoValueF{#2}{%
        \IfSubStr{#2}{m}{Gemini}{}%
        \IfSubStr{#2}{v}{\IfStrEq{#2}{v}{}{~}1.5 Pro}{}%
        \IfSubStr{#2}{e}{\IfStrEq{#2}{e}{Google}{~(Google)}}{}%
        \IfStrEq{#2}{p}{USA}{}%
        \IfStrEq{#2}{l}{Private}{}%
        \IfStrEq{#2}{S}{63,0}{}
        \IfStrEq{#2}{P}{96,5}{}
        \IfStrEq{#2}{C}{70,0}{}
        \IfStrEq{#2}{M}{22,6}{}
        \IfStrEq{#2}{Q}{42,8}{}
        \IfStrEq{#2}{TT}{257}{}
        \IfStrEq{#2}{FT}{9}{}
        \IfStrEq{#2}{TC}{0,19}{}
        \IfStrEq{#2}{FTPL}{\red{3}}{}
        \IfStrEq{#2}{FTMI}{\red{3}}{}
        \IfStrEq{#2}{FTAU}{0}{}
        \IfStrEq{#2}{FTPR}{\red{1}}{}
        \IfStrEq{#2}{FTRC}{\red{2}}{}
        \IfStrEq{#2}{FTJD}{0}{}
        \IfStrEq{#2}{PFT}{3,5}{}
        \IfStrEq{#2}{DS}{-7,9}{}
        \IfStrEq{#2}{DP}{-3,5}{}
        \IfStrEq{#2}{DC}{-1,7}{}
        \IfStrEq{#2}{DM}{-18,2}{}
      }%
    }%
    {s1}{\llm{claude}[\IfNoValueTF{#2}{}{#2}]}
    {s2}{\llm{deepseek}[\IfNoValueTF{#2}{}{#2}]}
    {s3}{\llm{qwen}[\IfNoValueTF{#2}{}{#2}]}
    {s4}{\llm{sabia}[\IfNoValueTF{#2}{}{#2}]}
    {s5}{\llm{llama}[\IfNoValueTF{#2}{}{#2}]}
    {s6}{\llm{gpt}[\IfNoValueTF{#2}{}{#2}]}
    {s7}{\llm{gemini}[\IfNoValueTF{#2}{}{#2}]}
    {s8}{\llm{mistral}[\IfNoValueTF{#2}{}{#2}]}%
    {p1}{\llm{claude}[\IfNoValueTF{#2}{}{#2}]}
    {p2}{\llm{mistral}[\IfNoValueTF{#2}{}{#2}]}
    {p3}{\llm{gpt}[\IfNoValueTF{#2}{}{#2}]}
    {p4}{\llm{sabia}[\IfNoValueTF{#2}{}{#2}]}
    {p5}{\llm{llama}[\IfNoValueTF{#2}{}{#2}]}
    {p6}{\llm{deepseek}[\IfNoValueTF{#2}{}{#2}]}
    {p7}{\llm{gemini}[\IfNoValueTF{#2}{}{#2}]}
    {p8}{\llm{qwen}[\IfNoValueTF{#2}{}{#2}]}%
    {c1}{\llm{claude}[\IfNoValueTF{#2}{}{#2}]}
    {c2}{\llm{gemini}[\IfNoValueTF{#2}{}{#2}]}
    {c3}{\llm{qwen}[\IfNoValueTF{#2}{}{#2}]}
    {c4}{\llm{deepseek}[\IfNoValueTF{#2}{}{#2}]}
    {c5}{\llm{llama}[\IfNoValueTF{#2}{}{#2}]}
    {c6}{\llm{sabia}[\IfNoValueTF{#2}{}{#2}]}
    {c7}{\llm{mistral}[\IfNoValueTF{#2}{}{#2}]}
    {c8}{\llm{gpt}[\IfNoValueTF{#2}{}{#2}]}%
    {m1}{\llm{claude}[\IfNoValueTF{#2}{}{#2}]}
    {m2}{\llm{deepseek}[\IfNoValueTF{#2}{}{#2}]}
    {m3}{\llm{sabia}[\IfNoValueTF{#2}{}{#2}]}
    {m4}{\llm{qwen}[\IfNoValueTF{#2}{}{#2}]}
    {m5}{\llm{gpt}[\IfNoValueTF{#2}{}{#2}]}
    {m6}{\llm{llama}[\IfNoValueTF{#2}{}{#2}]}
    {m7}{\llm{mistral}[\IfNoValueTF{#2}{}{#2}]}
    {m8}{\llm{gemini}[\IfNoValueTF{#2}{}{#2}]}%
  }[\textbf{ERRO: Modelo '#1' não encontrado}]%
}

\NewDocumentCommand{\allLLMs}{o}{%
  \llm{gpt}[\IfNoValueTF{#1}{}{#1}], 
  \llm{llama}[\IfNoValueTF{#1}{}{#1}], 
  \llm{claude}[\IfNoValueTF{#1}{}{#1}], 
  \llm{gemini}[\IfNoValueTF{#1}{}{#1}], 
  \llm{deepseek}[\IfNoValueTF{#1}{}{#1}], 
  \llm{mistral}[\IfNoValueTF{#1}{}{#1}], 
  \llm{qwen}[\IfNoValueTF{#1}{}{#1}], 
  and \llm{sabia}[\IfNoValueTF{#1}{}{#1}]%
}

\NewDocumentCommand{\allBestLLMs}{o}{%
  \llm{s1}[\IfNoValueTF{#1}{}{#1}], 
  \llm{s2}[\IfNoValueTF{#1}{}{#1}], 
  \llm{s3}[\IfNoValueTF{#1}{}{#1}], 
  and \llm{s4}[\IfNoValueTF{#1}{}{#1}]%
}

\NewDocumentCommand{\bestLLM}{o}{%
  \llm{s1}[\IfNoValueTF{#1}{}{#1}]
}

\NewDocumentCommand{\otherBestLLM}{o}{%
  \llm{s2}[\IfNoValueTF{#1}{}{#1}], 
  \llm{s3}[\IfNoValueTF{#1}{}{#1}], 
  and \llm{s4}[\IfNoValueTF{#1}{}{#1}]%
}

\NewDocumentCommand{\worstLLMs}{o}{%
  \llm{s8}[\IfNoValueTF{#1}{}{#1}], 
  \llm{s7}[\IfNoValueTF{#1}{}{#1}], 
  \llm{s6}[\IfNoValueTF{#1}{}{#1}], 
  and \llm{s5}[\IfNoValueTF{#1}{}{#1}]
}

\newcommand{\minFinalScore}{62,2}
\newcommand{\maxFinalScore}{70,9}
\newcommand{\diffMinMaxFinalScore}{8,7}
\newcommand{\diffMinMaxSuccessRate}{4,5}
\newcommand{\diffMaxAllMetricsBetweenFirstAndOtherBestLLMs}{7,5}
\newcommand{\minSuccessRateForBestLLMs}{95,5}
\newcommand{\minMutationForBestLLMs}{64,3}
\newcommand{\minCoverageForBestLLMs}{33,3}
\newcommand{\totalTests}{1.635}
\newcommand{\totalFailedTests}{39}
\newcommand{\totalFailedTestsRate}{2,38}

\newcommand{\RQOne}{%
    Are LLMs effective in generating integration tests that reflect the intended business logic and context? 
}

\newcommand{\RQTwo}{%
    Which LLM is the most effective in generating integration tests?
}

\title{Combining TSL and LLM to Automate REST API Testing: A Comparative Study}

\author{Thiago Barradas}
\affiliation{%
\institution{Universidade Federal Fluminense}
\city{Niterói}
\state{RJ}
\country{Brazil}}
\email{thbarradas@id.uff.br}

\author{Aline Paes}
\affiliation{%
\institution{Universidade Federal Fluminense}
\city{Niterói}
\state{RJ}
\country{Brazil}}
\email{alinepaes@ic.uff.br}  

\author{Vânia de Oliveira Neves}
\affiliation{%
\institution{Universidade Federal Fluminense}
\city{Niterói}
\state{RJ}
\country{Brazil}}
\email{vania@ic.uff.br}    


\renewcommand{\shortauthors}{Barradas et al.}


\begin{abstract}

The effective execution of tests for REST APIs remains a considerable challenge for development teams, driven by the inherent complexity of distributed systems, the multitude of possible scenarios, and the limited time available for test design. Exhaustive testing of all input combinations is impractical, often resulting in undetected failures, high manual effort, and limited test coverage. To address these issues, we introduce RestTSLLM, an approach that uses Test Specification Language (TSL) in conjunction with Large Language Models (LLMs) to automate the generation of test cases for REST APIs. The approach targets two core challenges: the creation of test scenarios and the definition of appropriate input data. The proposed solution integrates prompt engineering techniques with an automated pipeline to evaluate various LLMs on their ability to generate tests from OpenAPI specifications. The evaluation focused on metrics such as success rate, test coverage, and mutation score, enabling a systematic comparison of model performance. The results indicate that the best-performing LLMs -- \allBestLLMs[mve] --  consistently produced robust and contextually coherent REST API tests. Among them, \llm{claude}[mv] outperformed all other models across every metric, emerging in this study as the most suitable model for this task. These findings highlight the potential of LLMs to automate the generation of tests based on API specifications.

\end{abstract}



\maketitle


\newcommand{\Introduction}{%
    I\MakeLowercase{ntroduction}
}
\section{\Introduction}
Software testing is an essential component of the system development lifecycle, playing a crucial role in ensuring the quality and reliability of systems~\cite{tutejaImportanceTesting2012, jindalImportanceTesting2016, anandImportanceTesting2019}. Over the years, the increasing complexity of computational systems and the consequences of undetected failures have highlighted the need for robust and well-structured testing practices at all stages of development \cite{tutejaImportanceTesting2012, jindalImportanceTesting2016, anandImportanceTesting2019}.

Despite its importance, the effective execution of tests is not trivial. Teams often face challenges due to the complexity of systems, the volume of tests, and the limited time available for testing design and execution~\cite{tutejaImportanceTesting2012, newmanIntegrationTest2015, mendozaRelatedWork2024}. Exhaustively testing all possible inputs is unfeasible, especially in systems that require high reliability, which consumes a significant portion of the development effort \cite{tutejaImportanceTesting2012}. Additionally, manual test execution can result in incomplete scenarios and difficulty in detecting subtle failures \cite{tutejaImportanceTesting2012}, leading to high costs and limited coverage \cite{jindalImportanceTesting2016}.

Several studies have contributed to advancing software testing techniques, particularly by addressing challenges such as test input generation and the complexities inherent in real-world, modern systems~\cite{orsoPreviousStudies2014}. These challenges were highlighted in the 2014 study by Orso et al. \cite{orsoPreviousStudies2014} and remain open problems nowadays. Much progress is still needed to fully address all these difficulties, particularly in improving the effectiveness of generated tests (i.e., producing meaningful test scenarios) and automating testing processes end to end.

However, this process can be significantly accelerated by the recent emergence of several studies exploring the use of Large Language Models (LLMs) to enhance software testing techniques\cite{boukhlifStudyLLM2024}. LLMs have rapidly and consistently gained prominence in task automation and are increasingly recognized as promising tools to tackle many of the key challenges in software engineering in the coming years \cite{pezzePaperSE2030_2024}. These include reducing the costs associated with test case generation and execution, reducing the overall complexity of testing activities and improving the contextual relevance of automatically generated tests \cite{pezzePaperSE2030_2024, boukhlifStudyLLM2024}.

Modern and integrated system architectures widely adopt the REST API (Representational State Transfer, Application Programming Interface) model in software development due to its simplicity, flexibility, and scalability~\cite{lercherRESTApi2024}. The REST API has become the standard for communication between different services, providing a uniform and lightweight way to integrate heterogeneous systems through HTTP communication. One of the main advantages of REST is its compatibility with the web and its ability to handle large volumes of distributed transactions, where communication between independent components is essential to maintain modularity and scalability \cite{atlidakisRESTApi2019, lercherRESTApi2024}.

With the widespread adoption of REST APIs, integration testing is a crucial strategy for improving software quality because it ensures that different system components interact correctly, offering a more realistic view of how an application behaves in a production environment~\cite{sommerville2010, sotiriadis2017unitandintegration, golmohammadi2022testingrestfulapissurvey, newmanIntegrationTest2015}. Unlike unit tests, which target isolated functionalities, integration tests focus on interactions between services, modules, or systems, ensuring the correct exchange of information, such as inputs, responses, and communication flows. According to Newman \cite{newmanIntegrationTest2015}, failing to ensure proper integration between components can result in critical communication issues, such as inadequate responses. Corradini et al. \cite{corradinimIntegrationTest2021} further emphasize that for REST APIs, validating system behavior in real-world scenarios depends heavily on integration tests. 

Recent studies highlight the increasing attention to REST API testing, with emphasis on automation techniques, schema-based fuzzing, and test coverage metrics~\cite{golmohammadi2023TestingAPISurvey}. Golmohammadi et al.~\cite{golmohammadi2023TestingAPISurvey} identified black-box approaches and schema-guided fuzzers as the most common strategies, while also pointing out challenges like oracle definition and handling of real-world scenarios. Other works~\cite{corradinimIntegrationTest2021, tapplerInputData2024, olsthoornInputData2022} reinforce the importance of integration testing and the difficulties in generating valid inputs for complex interaction flows.

Creating such tests directly depends on the mapped scenarios and their respective input data. This process still faces several challenges, mainly due to the inherent complexity of interactions between system components, such as services or databases. A major obstacle lies in ensuring that inputs are both valid, and coherent with each component's constraints, which requires a deep understanding of business rules, and system contracts. Moreover, automatic input generation methods often fail to adequately exercise all critical integration paths \cite{olsthoornInputData2022, tapplerInputData2024, golmohammadi2023TestingAPISurvey}.

A recent survey by Wang et al. \cite{wangRelatedWork2024} reviews 102 studies on the application of LLMs in supporting various software testing tasks. In particular, LLMs have shown promise in generating test inputs, test oracles, and behavior-based tests--especially in contexts where textual descriptions are transformed into structured test cases--a key aspect in scenarios involving REST APIs, which often rely on standardized documentation such as OpenAPI\footnote{OpenAPI is a specification for describing REST APIs in a standardized, and machine-readable format.}. Despite the growing interest in applying LLMs to software testing, Wang et al. \cite{wangRelatedWork2024} also identify topics that remain underexplored. This is particularly true for more complex testing scenarios, such as those involving integrated systems and components.

Complementing the survey by Wang et al. \cite{wangRelatedWork2024}, the study by Alshahwan et al. \cite{alshahwanRelatedWork2023} extends these findings from an industrial perspective, emphasizing the importance of advancements in test automation, especially in complex environments such as the integration of distributed systems and REST APIs. Their study highlights challenges in test data generation, maintenance cost reduction, and improved test coverage--areas that can benefit from using artificial intelligence techniques such as LLMs. These observations reinforce the need for developing approaches that enhance the robustness and effectiveness of integration testing for REST APIs.

In this context, the main objective of this work is to introduce RestTSLLM, an approach for the automatic generation of integration tests for REST APIs based on their OpenAPI specification and using LLMs. To simplify the understanding by LLMs by decomposing the problem \cite{khot2023decomposed}, we suggest the use of an intermediate language, capable of structuring business specifications into structured test scenarios that can address all flows, validating both happy paths and edge cases, including validation failures. For this purpose, we propose the use of Test Specification Language (TSL) \cite{ostrand_TSL_CategoryPartition1988}, which serves as a bridge between business requirements and automated testing by providing a high-level, declarative format for specifying test cases. Instead of writing low-level test scripts, users define inputs, expected behaviors, and outcomes in a structured, human-readable format. By abstracting the test logic from its implementation, TSL enables the automatic generation of test scripts for different platforms, making it particularly effective for validating APIs and complex systems through reusable and maintainable specifications. Additionally, this study aims to evaluate the effectiveness of LLMs in understanding the context of these test scenarios and generating the corresponding set of integration tests. Thus, this research seeks to answer the following research questions (RQ):

\begin{itemize}
\renewcommand{\labelitemi}{$\textbf{RQ1:}$}
\item \RQOne
\end{itemize}

\begin{itemize}
\renewcommand{\labelitemi}{$\textbf{RQ2:}$}
\item \RQTwo
\end{itemize}

To answer these questions, this study adopts prompt engineering with the \textit{few-shot} and \textit{decomposed prompting} techniques \cite{khot2023decomposed}, teaching the LLM, part by part, how to produce appropriate responses to the prompts submitted to the model. Our approach includes an intermediate step for generating integration tests. First, we convert an OpenAPI specification into test cases using the TSL. Then, we convert the TSL test cases into functional integration tests using xUnit (.NET) \cite{xunit2025}. In presenting examples to the model, we demonstrate to the LLM how to follow these steps. The experiment also aims to execute the prompts across multiple projects and models previously selected based on criteria established in this study. After processing all prompts and generating the integration tests, we execute the tests produced by the models and collect and evaluate performance metrics, allowing us to answer the proposed research questions.

As a result of this experiment applied to six open-source projects, the models \allBestLLMs[mve] achieved the best results, and met the study's expectations, proving to be promising tools for this purpose. These tools enabled us to generate more efficient integration tests than the other evaluated LLMs. \llm{claude}[mv] outperformed all other models across every metric, emerging as the most suitable model for this task. The models \worstLLMs[mve] performed lowest but were still considered satisfactory.

This work presents as its main contributions the proposed approach, RestTSLLM, for integration test generation using TSL and LLMs; the performance comparison of different LLMs in generating integration tests according to our method, identifying those with the best results; and the development of an automated script capable of integrating multiple LLMs, which also allows easy connection to additional models for future research and reuse of the code, enabling a wide range of experiments in a faster and more adaptable manner across different usage contexts.

The data and code supporting the conclusions of this study are publicly available on GitHub \cite{comparativeLLMsIntegrationTestGithub2025}.

The RestTSLLM approach aims to support researchers, developers, and companies in improving REST API testing. Automating test generation can reduce time, cost and effort while enhancing fault detection. The flexibility of LLMs also facilitates faster adaptation to evolving requirements. Finally, this study fills a gap in the literature by evaluating the use and performance of LLMs in REST API testing.

The remainder of this paper is organized as follows: Section~\ref{sec:relatedworks} reviews related work. Section~\ref{sec:approach} presents more details about our approach, RestTSLLM, and how it can be replicated. Section~\ref{sec:methodology} describes the methodology used, including details of the selected technologies and tools, the prompts utilized, and the evaluation and comparison metrics used in this study. Section~\ref{sec:results} delves into analyzing the results obtained in the experiment. Section~\ref{sec:limitations} discusses threats to the validity and limitations of the study regarding its applicability. Finally, Section~\ref{sec:conclusion} presents the conclusions, summarizing the main contributions and outlining possible future work.

\newcommand{\RelatedWorks}{%
    R\MakeLowercase{elated works}
}
\section{\RelatedWorks}\label{sec:relatedworks}

This section reviews relevant studies that explore the use of Large Language Models (LLMs) in software engineering, with a particular focus on test automation and REST API testing, and provides a summary of the identified research gaps that motivate our study.

There is a growing interest in using LLMs across various areas of software engineering, including requirement engineering, code implementation, testing, maintenance, and deployment \cite{hou2024large, fanLLMImportance2023}. Fan et al. \cite{fanLLMImportance2023} provide a comprehensive overview of this trend, highlighting the diverse applications of LLM tools in tasks such as requirements, coding, bug fixing, refactoring, performance improvement, design, documentation, and analysis. Moreover, the study points out key technical challenges that come with these advances, particularly the need for reliable methods to detect and correct incorrect solutions.

The study by Belzner et al. \cite{belznerRelatedWork2023} provides an insightful overview of the benefits and challenges of using LLMs in software engineering, focusing on specific stages of the development cycle such as requirements engineering, system design, code generation, and testing. The results indicate that LLMs--GPT 3.5 and Bard (now Gemini)-- can effectively assist in creating helpful software engineering artifacts, particularly during simpler phases of development. However, it presents limitations in terms of scalability and accuracy, especially in more complex or ill-defined tasks such as integration testing. Although the approach analyzes various development lifecycle phases, it provides limited attention to integration testing and complex testing scenarios.

In comparative performance studies, Mendoza et al. \cite{mendozaRelatedWork2024} conducted a comparative analysis of different LLM tools, including GPT 3.5, GPT 4, Copilot, and Gemini, to assess their ability to generate input data from BDD scenarios. The study highlighted that, although some tools struggled to understand certain contexts, GPT 4 and Gemini performed better in generating test data consistent with BDD specifications, outperforming GPT 3.5 and Copilot. The analysis reinforces the importance of adapting LLM usage to improve testing efficiency, particularly in more complex and dynamic scenarios.

The work by Ouédraogo et al. \cite{ouedraogoRelatedWork2024} presents a comprehensive analysis of LLMs in the generation of unit tests, exploring the effectiveness of different models such as GPT 3.5, GPT 4, Mixtral 7B, and Mixtral 8x7B, along with advanced prompt engineering techniques, including  \textit{zero-shot}, \textit{few-shot}, \textit{chain-of-thoughts}, and \textit{tree-of-thoughts}. The results indicated that GPT 3.5 achieved the best performance. Furthermore, it revealed that although LLMs show significant potential, there are still limitations related to the quality of the generated tests and the presence of test smells that may compromise long-term maintainability. Even though the study focuses on unit testing—which is inherently simpler and requires less contextual information than more complex tasks like integration testing—it highlights the need to evolve test generation methods and refine prompting strategies to improve test coverage and clarity, especially when comparing LLMs to traditional tools.

Kim et al. \cite{kimArticleLeveragingLLM2024} conducted an experiment on using LLMs to improve REST APIs testing and proposed the RESTGPT approach. The paper mentions that traditional test generation tools for REST APIs typically use an OpenAPI specification as input. However, the tools cannot understand the insights provided due to the lack of rules exposed in a machine-readable way. To address this problem, RESTGPT proposes using an LLM to enrich the information provided in the specification. It takes an OpenAPI file as input and generates constraints, rules, and examples, returning a more detailed and machine-readable OpenAPI specification. This enhanced output can then be used by other automated test generation tools as input. This method showed significant improvements in the accuracy of correctly interpreting parameter descriptions and generating valid values more accurately and contextually than traditional tools such as NLP2REST and ARTE. The study reports notable gains in rule extraction accuracy. However, this work does not aim to generate automated REST API tests, but rather to enrich the raw material used as input for this task.

Expanding upon this line of research, Kim et al.~\cite{kim2025llamaresttesteffectiverestapi} proposed LlamaRestTest, a black-box testing approach based on fine-tuned and quantized LLaMA 3 models to improve input generation and detection of parameter dependencies in REST API testing. Unlike RESTGPT, which statically enriches OpenAPI specifications, LlamaRestTest dynamically adapts test inputs based on server feedback using reinforcement learning. It employs two specialized models: one for identifying dependencies (LlamaREST-IPD) and another for generating valid inputs (LlamaREST-EX). The study shows that small, optimized models can outperform larger ones and existing tools in coverage and fault detection. However, it focuses on dynamic test execution rather than generating reusable test artifacts, making each run unique and harder to reproduce or version.

In contrast to LlamaRestTest, our approach uses a simpler and more accessible workflow that generates reusable test artifacts covering both success and client error scenarios. We use general-purpose LLMs, without the need for prior training, to generate REST API tests only based on examples provided through prompt engineering. We explore the direct use of LLMs in their default state to produce executable tests compatible with traditional tools, promoting easier adoption and integration into developers' daily workflows.

In addition to the previously mentioned approaches that leverage LLMs, studies such as those by Golmohammadi et al.~\cite{golmohammadi2023TestingAPISurvey}, Viglianisi et al. \cite{viglianisi2020resttestgen}, and Corradini et al. \cite{corradini2021empirical} analyze and explore the automated generation of REST API tests from OpenAPI specifications using traditional black-box tools, including RESTTestGen, RESTler, bBOXRT, and RESTest. While these tools have shown satisfactory coverage and fault detection results, they still face challenges in balancing these two metrics. It is worth noting that such studies do not explore the use of LLMs to optimize outcomes or to enhance the interpretation of natural language rules present in the specifications—limiting their potential for understanding complex business logic.

Despite the increasing number of studies on using LLMs for test generation, which demonstrate their potential, there remains a lack of research focused on generating and evaluating REST API tests. Most approaches concentrate on unit testing, with a limited focus on integration tests, and the few that exist for integration testing do not have an approach capable of solving most problems efficiently. In this context, we identify an opportunity to expand research and fill this gap regarding the automated generation of REST API tests, providing a new approach to automating them, and enabling comparative evaluation of effectiveness among existing LLMs, providing a broader understanding of how LLMs can be used to automate the production of integration tests for REST APIs.

\newcommand{\OurApproach}{%
    O\MakeLowercase{ur approach}
}
\section{\OurApproach}\label{sec:approach}

\begin{figure}[hb]
  \centering
  \includegraphics[width=.48\textwidth]{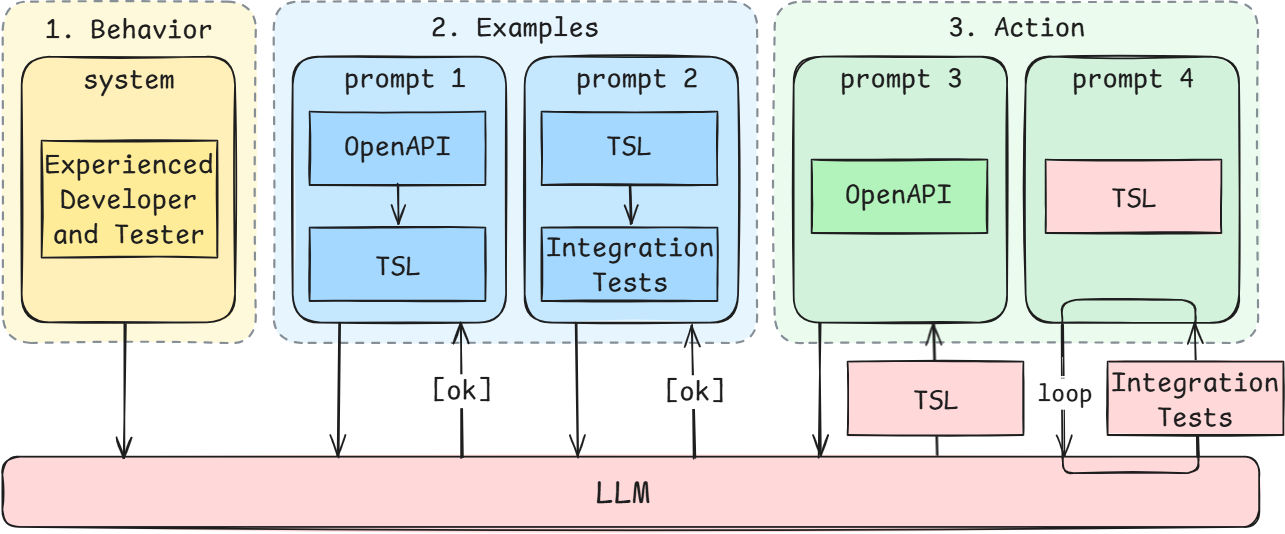}
  \caption{Overview of RestTSLLM approach}
  \Description{}
  \label{fig:approach_RestTSLLM}
\end{figure}

RestTSLLM is an approach designed to address the challenges of REST API testing, particularly in generating scenarios, input data, and executable tests. The method is based on the premise of being easily replicable, without requiring fine-tuning or specialized models, relying solely on prompt engineering and the use of the \textit{few-shot} and \textit{decomposed prompting} techniques. Another essential premise is that its structure is easily adaptable and extensible to other types of tests that face similar challenges and other technologies. This section aims to detail our approach, which is visually summarized in Figure~\ref{fig:approach_RestTSLLM}.
 


\vspace{-0.5em}
The proposed prompt engineering is organized into three main stages: (1) Behavior -- defining the behavior of the LLM, (2) Examples -- presenting examples, and finally, (3) Action -- requesting the execution of the task.

The diagram in Figure~\ref{fig:approach_RestTSLLM} adopts a color-coded structure to distinguish different components and roles within the RestTSLLM approach visually. The three main stages are highlighted in distinct colors: yellow for Behavior, blue for Examples, and green for Action. The LLM and all outputs generated by it are shown in red. It is important to note that in \textit{prompt 4}, the TSL input is shown in red, to indicate that it is composed of the output previously generated by the LLM in \textit{prompt 3}.

In the first stage, \textbf{Behavior}, the LLM is instructed, through the \textit{system} parameter, to act as an experienced developer and tester, capable of understanding REST API specifications and generating coherent and functional integration test artifacts. This sets the expected behavior for the model in the following stages. It also allows for restricting or enabling functions, guiding the conversational flow, imposing ethical or safety constraints, and preparing the response environment, among other possibilities that limit or direct its behavior. The use of this technique allows us to improve the performance and generalization of the generated result, ensuring greater alignment with the expectations of how the generation of results will be done given the established context \cite{mu2025closerSystemPrompt, zhang2024sprigSystemPrompt, nebulySystemPrompt2025}.

In addition to acting as an experienced developer and tester, the model was instructed to operate as if it had a strong knowledge of the target technology used for test generation, of rule extraction from OpenAPI and conversion to TSL using the Category-Partition method \cite{ostrand_TSL_CategoryPartition1988}, and of testing best practices such as organizing tests in the AAA pattern (Arrange, Act, Assert), boundary testing, equivalence classes, and scenario design aiming for high coverage. Furthermore, the model's behavior was configured to return only the requested code, without textual explanations or justifications.

The second stage, \textbf{Examples}, consists of two steps that apply the \textit{few-shot} and \textit{decomposed prompting} techniques \cite{khot2023decomposed} to guide the model's learning through concrete examples: \textbf{Prompt 1:} An OpenAPI specification is presented as input, and the desired output is a set of test cases written in Test Specification Language (TSL), whose main objective is to provide a formal, and standardized language for the specification, and documentation of software tests, usually formatted in \textit{yml}. TSL is designed to help define structured test cases, including scenarios, inputs, expected outputs, and conditions, in a clear, and understandable way, promoting greater accuracy, reusability, and automation in the testing process \cite{ostrand_TSL_CategoryPartition1988}. This stage teaches the LLM how to interpret OpenAPI documents and extract relevant information to define structured test scenarios; \textbf{Prompt 2:} The TSL presented in the first step is now used as input, and the expected output is a set of executable integration tests. The approach supports the generation of tests in any technology—depending on what is presented in the model. This step teaches the LLM how to translate abstract test cases into functional code based on the structure and conventions of the chosen technology while following integration testing best practices.

This intermediate generation step using TSL was designed to improve the LLM’s effectiveness. By separating the test case reasoning from the implementation, the first prompt allows the model to concentrate exclusively on understanding the business rules described in the API specification, without being burdened by concerns related to code structure or syntax. Once the test scenarios are clearly defined in TSL, the second prompt focuses solely on converting those predefined cases into functional test code. This separation of concerns helps the LLM perform more efficiently in each task, reducing complexity and improving the output quality \cite{khot2023decomposed}.

The third stage, \textbf{Action}, also contains two steps in which the LLM applies the logic learned from the examples to new inputs. Here, the model generates the outputs: \textbf{Prompt 3:} A real OpenAPI specification—different from the one used in the examples—is provided to the LLM. Based on what it learned in \textit{prompt 1}, the model generates a new TSL, describing the scenarios that should be tested; \textbf{Prompt 4:} The TSL generated in the previous step is then used as input. Following the logic of \textit{prompt 2}, the model produces the final integration test code. These tests are ready to be executed according to the framework and language defined in the example prompts provided to the LLM. 

During the experiments, we observed that some LLMs — particularly those with lower output token limits—had difficulty generating all the tests in a single response. This often resulted in truncated outputs. To address this, we implemented a strategy where \textit{prompt 4} is executed in a loop, requesting the generation of subsets of tests grouped by specific tags or methods defined in the OpenAPI specification. These tags are carried over to the TSL to segment the test cases, allowing the LLM to produce valid output in manageable parts while preserving the overall structure and completeness of the test suite.

Listing~\ref{lst:tsl-sample} and Listing~\ref{lst:code-sample} illustrate the outputs generated by the LLM in response to prompts 3 and 4, respectively. The first block shows a test case written in TSL, which defines the scenario “Login Valid Credentials Returns Token” in a structured and declarative format. This format allows the model to describe the endpoint to be tested, required preconditions, input data, and expected output. Listing~\ref{lst:code-sample} presents the corresponding integration test code generated from the TSL using xUnit (.NET). It includes all the necessary steps to perform the test according to the defined expectations. In general, the input data defined in the TSL is preserved during the test generation, although minor adjustments may be applied to ensure repeatability. In the example show in Listing~\ref{lst:code-sample}, for instance, a function was used to dynamically generate a unique email address, replacing the static value originally defined in the TSL. This strategy prevents failures in subsequent test executions that could occur due to uniqueness constraints in the system under test, as taught in the examples presented to LLM.

\begin{listing}[ht]
\caption{Sample of TSL generated by LLM}
\label{lst:tsl-sample}
\begin{minted}[fontsize=\scriptsize, breaklines]{yaml}
- id: TC101
  group: Account
  name: Login Valid Credentials Returns Token
  endpoint: /api/accounts/tokens
  method: POST
  preconditions:
    - "User with email 'valid@test.com' exists"
  request_body:
    email: "valid@test.com"
    password: "Val1d!Pass"
  expected_response:
    status_code: 200
    body:
      userId: is string not empty
      token: is string not empty
      refreshToken: is string not empty
\end{minted}
\end{listing}

\begin{listing}[ht]
\caption{Sample of integration test generated\\by LLM from previous TSL}
\label{lst:code-sample}
\begin{minted}[fontsize=\scriptsize, breaklines]{csharp}
[Fact] 
public async Task TC101_Login_Valid_Credentials_Returns_Token()
{
    // Arrange
    var email = GenerateUniqueEmail();
    var password = "Val1d!Pass";

    // Create user first
    await CreateUserAsync(new
    {
        firstName = "John",
        lastName = "Doe",
        email,
        password,
        isAdmin = false
    });

    // Act
    var response = await LoginAsync(email, password);

    // Assert
    var body = await response.Content.ReadFromJsonAsync<JsonObject>();
    Assert.Equal(HttpStatusCode.OK, response.StatusCode);
    Assert.False(string.IsNullOrEmpty(body["userId"].ToString()));
    Assert.False(string.IsNullOrEmpty(body["token"].ToString()));
    Assert.False(string.IsNullOrEmpty(body["refreshToken"].ToString()));
}
\end{minted}
\end{listing}

\newcommand{\ExperimentalMethodology}{%
    E\MakeLowercase{xperimental methodology}
}
\section{\ExperimentalMethodology}\label{sec:methodology}

\begin{figure}[ht]
  \centering
  \includegraphics[width=.48\textwidth]{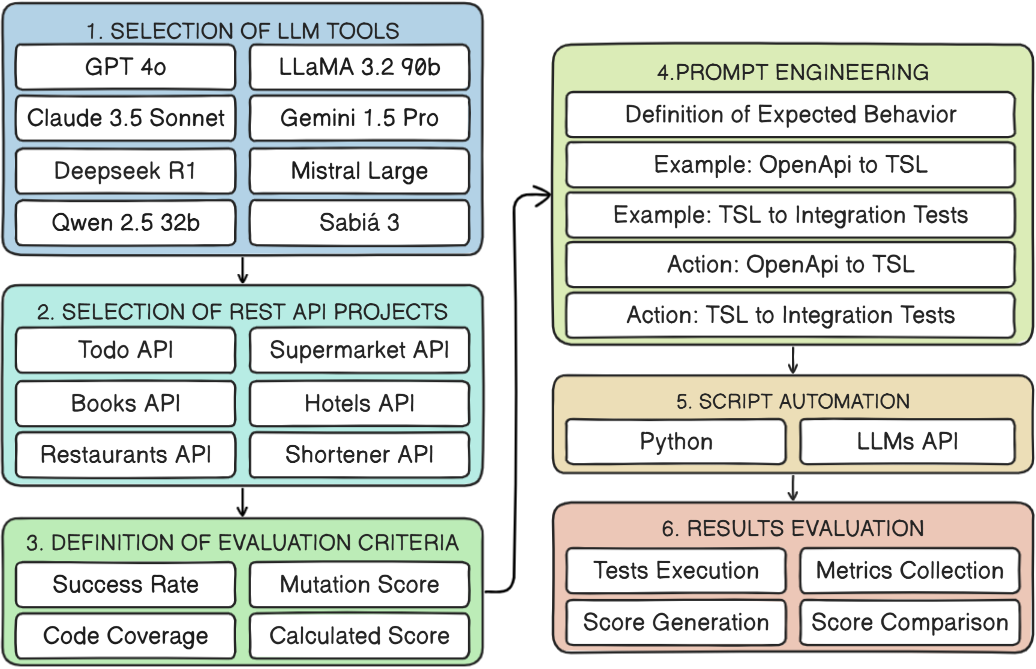}
  \caption{Experimental methodology flow}
  \Description{}
  \label{fig:methodology}
\end{figure}

This section describes the methodology used to conduct the experiment. Through these steps, we aim to validate the proposed approach by evaluating the effectiveness of the selected LLMs to determine which one best aligns with our expectations for integration test generation. Figure~\ref{fig:methodology} illustrates the methodological steps. Each step is summarized below, with further details presented in the following subsections.

The first step, \textbf{Selection of LLM tools}, involves choosing a diverse set of language models to support the evaluation of our proposed approach. We selected \allLLMs[mve], which stand out in popularity, and performance in code generation tasks. We also included \llm{sabia}[mve], a Brazilian LLM designed to support Portuguese (Subsection~\ref{subsec:llmtools}). The second step,  \textbf{Selection of REST API Projects}, consisted of selecting six REST API repositories that met the following criteria: authorized for use in this study, contained an OpenAPI specification, were relevant on GitHub, had at least one integration with a database or external service, and developed using the target technology of the experiment (Subsection~\ref{subsec:projectselection}). 

The third step, \textbf{Definition of Evaluation Criteria}, involved defining the set of metrics to assess the effectiveness of the integration tests generated by the LLM tools. For this purpose, we select the success rate, coverage, and mutation score to compute a calculated score from these metrics and evaluate the performance of LLMs (Subsection~\ref{subsec:evaluationcriteria}). The fourth step, \textbf{Prompt Engineering}, expands on the experiment details in creating prompts used to guide the LLMs in RestTSLLM approach (Subsection~\ref{subsec:promptengineering}). The fifth step, \textbf{Script Automation}, refers to the development of an automated script for integrating with the LLMs to process the prompts, and their results (Subsection~\ref{subsec:scriptautomation}). Finally, the sixth step,  \textbf{Results Evaluation}, describes the evaluation of the integration tests generated by the LLMs (Subsection~\ref{subsec:resultsevaluation}).

\subsection{Selection of LLM tools}\label{subsec:llmtools}

As mentioned earlier, the LLM tools selected for this study include \allLLMs[mve]. The selection criteria were based on the popularity of LLMs, extracted from publicly available online benchmarks \cite{openrouterLLMSelection2025, vellumLLMSelection2025, uniteLLMSelection2025, definitionLLMSelection2025, livebenchLLMSelection2024, shakudoLLMSelection2025, botpressLLMSelection2025}, and their use in previous studies \cite{xiaLLMSelection2024, aydinLLMSelection2025, qiuLLMSelection2024, hagosLLMSelection2024}. Additionally, only models with publicly available documentation, and the ability to be used both via API, and user interface were considered. 

\newcommand{\llmsTableContent}{} 
\newcommand{\llmsTableLine}[1]{%
  \llm{#1}[m] & \llm{#1}[v] & \llm{#1}[e] & \llm{#1}[p] & \llm{#1}[l] \\
}

\makeatletter
\foreach \model in \llmsEnum {
  \edef\temp{\noexpand\g@addto@macro\noexpand\llmsTableContent{%
    \noexpand\llmsTableLine{\model}%
  }}
  \temp
}
\makeatother

\begin{table}[ht]
  \caption{Selected LLMs}
  \label{tab:llms}
  \begin{tabular}{lllll}
    \toprule
    Model&Version&Company&Country&License\\
    \midrule
    \llmsTableContent
    \bottomrule
\end{tabular}
\end{table}

From this list, we selected the seven most popular models. In addition to these, we included one more LLM to ensure the presence of a model with higher accuracy in interpreting prompts in Portuguese—the main language in the prompts in this study. We chose the Brazilian model \llm{sabia}[mve] \cite{abonizioLLMSelectionMaritaca2024}, developed in the authors' home country, to support analyses in the local language, contributing to technological advancement, and highlight Brazilian research within the scientific community.

The final list of the selected LLMs is available in Table~\ref{tab:llms}, which indicates the selected models, the version of each model used (the most recent version available at the time of selection was considered), the provider company, its origin country, and whether the model is available as open-source or only through paid versions. 

\subsection{Selection of REST API projects}\label{subsec:projectselection}

The criteria used for selecting REST API projects for this experiment include: having a formal or permissive license that authorizes free use in this study; containing at least one REST API application, which is the focus of the research; providing an OpenAPI specification to serve as a basis for generating integration tests; demonstrating relevance on GitHub, with a combined number of stars and forks greater than 100, to avoid non-functional or unvalidated projects; including at least one integration with a database or another API to ensure some level of integration testing; offering documentation with installation, configuration, and usage guides to support comprehension and reproducibility; having a limited number of endpoints to constrain the resources required for the experiment; and being primarily implemented in recent versions of .NET—a technology supported by a leading software company (Microsoft), deeply mastered by the authors, and widely adopted in the industry for REST API development \cite{fortune_dotnet_2023}.

After searching on GitHub for projects that met the defined criteria, we selected six .NET-based REST API projects with at least one integration, an OpenAPI specification, and documentation in a \texttt{README.md} file. These projects and the evaluation criteria are presented in Table~\ref{tab:projects}. Most had an MIT license\footnote{\url{https://opensource.org/license/mit}}, allowing free use, while formal permission was obtained for the \textit{hotels-api} project\footnote{Authorization formally requested: \url{https://github.com/uffsoftwaretesting/RestTSLLM/tree/main/pdfs/use_permission_hotel_api_.pdf}}. We also included a \textit{CLOC}\footnote{CLOC stands for "Count Lines Of Code". We used the project available at \url{https://github.com/AlDanial/cloc} to count only C\# lines.} column to indicate the size of each project based on the number of lines of C\# code.

\begin{table*}[h]
  \caption{Selected Projects}
  \label{tab:projects}
  \begin{tabular*}{\textwidth}{@{\extracolsep{\fill}}llrrlrrrr}
    \toprule
      Name &
      License &
      .NET Version &
      Endpoints &
      Dependencies & 
      Stars & 
      Forks & 
      Stars + Forks &
      CLOC \\
    \midrule
     todo-api \cite{projectTodoApiGithub2025} &
        MIT & 7.0 & 7 & 
        SQLite & 
        2.964 & 441 & 3.405 & 1.576 \\
     supermarket-api \cite{projectSupermarketApiGithub2025} & 
        MIT & 8.0 & 8 &
        SQLite &
        484 & 166 & 550 & 977 \\
     books-api \cite{projectBooksApiGithub2025} & 
        MIT & 8.0 & 5 & 
        Postgre SQL, Redis & 
        145 & 92 & 237 & 809 \\
     hotels-api \cite{projectHotelsApiGithub2025} & 
         Requested & 
         7.0 & 14 & SQL Server & 
         76 & 42 & 118 & 2.843 \\
     restaurants-api \cite{projectRestaurantsApiGithub2025} &
         MIT & 8.0 & 10 & 
         SQL Server, Azure Storage & 
         64 & 39 & 103 & 1.804 \\
     shortener-api \cite{projectShortenerApiGithub2025} &
         MIT & 9.0 & 3 & 
         MongoDB, Redis &
         78 & 22 & 100 & 1.055 \\
  \bottomrule
\end{tabular*}
\end{table*}


\subsection{Definition of evaluation criteria}\label{subsec:evaluationcriteria}

We adopted test success rate, coverage, and mutation score as evaluation criteria. The success rate is an immediate indicator of the system stability for the executed set of integration tests, reflecting whether any failures were observed in the evaluated scenarios. However, the isolated analysis of this metric is insufficient to assess test quality, making it necessary to use complementary metrics \cite{jainMutationCoverage2023}.

Coverage is a widely used metric in software testing, frequently reported alongside the success rate. It quantifies the proportion of the software exercised during test execution, typically at the level of statements or branches. In this study, we focus on branch coverage, as it offers a more precise assessment of control-flow exploration by checking whether all decision outcomes are exercised. Higher coverage tends to indicate more comprehensive evaluation and may increase the likelihood of exposing failures within the system under test \cite{ouedraogoRelatedWork2024}.

The mutation score, in turn, is considered a stricter metric than coverage, although complementary to it. It assesses the sensitivity of a test suite to small changes in the system's logic by introducing artificial faults (mutants) and verifying whether the tests detect them. A high mutation score indicates that the tests effectively identify behavioral deviations, suggesting a stronger alignment with the intended functional requirements \cite{zhangMutationCoverage2022}.

Thus, the combination of these metrics was adopted as the evaluation criterion in this study, as it allows a comprehensive analysis of key aspects of software testing quality: ensuring that the system behaves as expected, guaranteeing a significant amount of coverage, and effectiveness of the test suite in detecting subtle logic faults\cite{jainMutationCoverage2023, zhangMutationCoverage2022}.

To compute a \textit{Calculated Score}, we applied the TOPSIS technique (Technique for Order Preference by Similarity to Ideal Solution \cite{madanchian_TOPSIS_MCDM_2023}), derived from the MCDM (Multi-Criteria Decision Making) method\cite{madanchian_TOPSIS_MCDM_2023}. TOPSIS aims to solve decision-making problems that involve multiple conflicting criteria in their analysis. These methods help select the best alternative in situations where different criteria, sometimes conflicting, must be considered simultaneously.

The formula used to calculate the \textit{Calculated Score}, denotated by \( S \), is presented below:
\vspace{0.1cm}
\[
S = w \cdot SuccessRate + w \cdot Coverage + w \cdot MutationScore
\]

\vspace{0.01cm}

The formula uses the previously mentioned metrics: success rate (\textit{SuccessRate}), coverage (\textit{Coverage}), and mutation score (\textit{MutationScore}). These metrics do not require normalization, as they are already expressed within a standard range between 0 and 1, represented as percentages. 
Finally, we applied a weight of 33.3$\overline{3}$\%, denotated by \( w \), to each metric to balance and proportionally reflect their relative importance in the calculated score regarding the effectiveness of the generated tests.

To compare the effectiveness of the LLMs, we used the average calculated score as the final score of their performance.

These evaluation criteria were designed to address part of the research questions proposed in this study. We combined quantitative and qualitative evaluation strategies to answer the proposed research questions fully. We performed a manual and subjective analysis for \textbf{RQ1}, which investigates whether LLMs can generate integration tests aligned with the intended logic and context. All test cases and generated code were reviewed by the first author to assess their coherence with the described business rules, the clarity of the scenarios, and their correspondence with the structure defined by the OpenAPI specification and the expected test patterns. The effectiveness aspect of \textbf{RQ1}, as well as the full assessment required by \textbf{RQ2}, was addressed through the objective metrics described above. The average calculated score derived from these metrics enabled us to rank the models according to their overall performance and identify which LLMs were the most efficient.

\subsection{Prompt engineering}\label{subsec:promptengineering}

In addition to the general aspects of the RestTSLLM approach previously explained in Section~\ref{sec:approach}, we will explain in this section the complementary peculiarities of applying the approach in our experimental methodology.

As the target technology for generating REST API integration tests, we chose .NET with xUnit \cite{xunit2025}, given the authors' depth and prior experience with the technology, supported by a leading software company (Microsoft), and widely adopted in the industry for developing and testing REST APIs \cite{fortune_dotnet_2023}. Given this choice, for our experimental evaluation, the specific examples presented in \textit{prompt 2}—converting TSL to integration tests—provided examples of tests written in .NET with xUnit. Consequently, in \textit{prompt 4}, the tests generated for the inputs of the previously selected real projects were also produced by the model in this technology.

The prompt content was written and executed in English and Portuguese while the OpenAPI specifications remained in English. We observed that the performance was equivalent and suitable when testing both fully-English and partially-Portuguese prompts, given that the smallest part of the prompt is textual instructions and the predominant content is examples and codes. We chose to keep the prompts in Portuguese, the native language of the authors' home country, aiming to support analyses focused on the local language, contribute to technological development, and highlight Brazilian research within the scientific community.

The generic prompts used in this study are available on GitHub\footnote{\url{https://github.com/uffsoftwaretesting/RestTSLLM/tree/main/general-files/default-prompts/files.md}}.

\subsection{Script automation}\label{subsec:scriptautomation}

Studies involving LLMs commonly apply prompt engineering in their experiments using the interface provided by the model developer \cite{wangRelatedWork2024}. However, this approach can become complex when executing the same sequence of prompts and collecting their results across a larger set of LLMs. For our study, we developed a Python script capable of integrating with the studied LLMs, allowing easy connection to additional models for future research, and reuse of the code.

To overcome this, we developed a Python script that automates the entire process: loading prompt files, handling model configurations, executing each step while maintaining conversational context, and saving the outputs and metrics. This made it possible to run consistent, large-scale experiments efficiently across different LLMs. All selected LLMs were accessed via remote APIs; none were executed locally.


The source code of the automated script is available on GitHub\footnote{\url{https://github.com/uffsoftwaretesting/RestTSLLM/tree/main/llm-processor/README.md}}.

\subsection{Results evaluation}\label{subsec:resultsevaluation}

The final execution that resulted in the outcomes of this study took place on March 9, 2025, and lasted 2 hours, and 30 minutes to process all prompts across all LLMs. The execution was done on a computer with an i7 processor and 32 GB RAM.
The results of the prompt executions in the LLMs can be accessed on GitHub\footnote{\label{fn:files}\url{https://github.com/uffsoftwaretesting/RestTSLLM/tree/main/projects/files.md}}.

After generating integration tests by LLMs in xUnit with .NET, the target project was duplicated for each LLM, the tests were manually copied, and then executed with the collection of the metrics described in Subsection~\ref{subsec:evaluationcriteria}. 

To obtain the success rate and coverage results, we ran the tests using the Visual Studio 2022 tool \cite{xunit2025}. We used the branch coverage metric previously explained in Subsection~\ref{subsec:evaluationcriteria}. 

To obtain the mutation score we use the Stryker.NET tool \cite{stryker2025} with maximum mutation level enabled (\textit{Advanced}), creating mutants for regex, string literals, collection initializer, statements like block, checked, and assignment, and operators like arithmetic, logical, bitwise, equality, boolean, update, and unary, and methods like string, math, and linq \cite{stryker2025}. We disabled the option to combine mutants in the same execution for greater assertiveness of the result, even though this option results in a longer execution time, given that mixed mutants can present unwanted side effects \cite{stryker2025}.

The metrics collected from the test executions, and the evidence of these executions are available in a public PDF file on GitHub\footnote{\label{fn:results}\url{https://github.com/uffsoftwaretesting/RestTSLLM/tree/main/pdfs/all_results.pdf}} due to space restrictions in this document.

\newcommand{\ResultsAndDiscussion}{%
    R\MakeLowercase{esults and discussion}
}
\section{\ResultsAndDiscussion}\label{sec:results}

This section presents and discusses the results obtained by applying the proposed RestTSLLM approach. The findings are organized according to the research questions (RQ) defined in this study.

To answer \textbf{RQ1} and \textbf{RQ2} we computed a final score (average calculated score) for each model, based on a weighted combination of success rate, coverage, and mutation score using the TOPSIS technique. The aggregated result, ordered from highest to lowest, are presented in Table~\ref{tab:results}. The detailed results by LLM and project are available in a public PDF file on our GitHub repository\hyperref[fn:results]{\textsuperscript{\ref{fn:results}}}. Although it is not part of the evaluation criteria, we included the number of tests generated and the total cost per generation to support our analysis. For simplification purposes, we use a single letter to represent the average of each metric:

\begin{itemize}
    \item \( S \) → Calculated Score  
    \item \( SR \) → Success Rate  
    \item \( C \) → Coverage
    \item \( M \) → Mutation Score
    \item \( T \) → Number of Tests 
    \item \( TC \) → Total Cost (USD)  
\end{itemize}

\newcommand{\resultsTableContent}{} 
\newcommand{\resultsTableLine}[1]{%
    \small{\llm{#1}[mv]}   & 
    \small{\llm{#1}[S]\%}  & 
    \small{\llm{#1}[P]\%}  & 
    \small{\llm{#1}[C]\%}  & 
    \small{\llm{#1}[M]\%}  & 
    \small{\llm{#1}[Q]}    & 
    \small{\$\llm{#1}[TC]} \\ 
}

\makeatletter
\foreach \model in \llmsScoreEnum {
  \edef\temp{\noexpand\g@addto@macro\noexpand\resultsTableContent{%
    \noexpand\resultsTableLine{\model}%
  }}
  \temp
}
\makeatother

\begin{table}[ht]
  \caption{Average of the metrics for the tests \\generated by LLMs}
  \label{tab:results}
  \begin{tabular}{lcccccc}
    \toprule
      \small{Model}             &
      \small{\( \textbf{S} \)}  &
      \small{\( SR \)}          &
      \small{\( C \)}           &
      \small{\( M \)}           &
      \small{\( T \)}           &
      \small{\( TC \)}          \\
    \midrule
    \resultsTableContent
    \bottomrule
\end{tabular}
\end{table}

\subsection{RQ1: \RQOne}\label{subsec:resultsRQ1}

From a qualitative perspective, the application of \textit{few-shot} and \textit{decomposed prompting} techniques—using examples with expected outputs in a consistent and functional format—proved to be satisfactory. All LLMs managed to preserve the expected structure and generate compilable code for integration test execution.

The content generated was read and manually reviewed. All results were satisfactory in terms of readability, clarity, and adherence to the test patterns presented in the prompts. The test cases were aligned with the business logic described in the OpenAPI specifications, demonstrating that the models understood the intended behavior of the APIs. In the manual review, we validated the alignment between the generated content and the business rules by analyzing details such as authentication criteria, status codes, input and output properties, as well as whether both success scenarios and relevant edge cases made sense in the context of the specification. Among the evaluated APIs, only \textit{todo-api} had pre-existing tests, which showed no structural or naming convention similarity with the outputs generated by the models. All generated test cases and integration code are available on GitHub\hyperref[fn:files]{\textsuperscript{\ref{fn:files}}}.

In addition to this subjective evaluation, the effectiveness of the tests generated by each model was measured using three key metrics: success rate, coverage, and mutation score. All models achieved average success rates above \llm{p8}[P]\%, with balanced coverage and mutation scores in most cases. Only \totalFailedTestsRate\% of all generated tests had some kind of failure, which we detail in Subsection~\ref{subsec:analysisfailures}. This indicates that LLMs are capable of producing tests that not only execute successfully but also explore the system under test meaningfully.

It is important to note that the tests had a relatively low mutation rate given the black-box test generation, which, by design, omits the implementation aspects of the test design, and may cause this side effect, indicating a possible gap between the specification and the implementation. Despite this, compared to the only project that had tests, \textit{todo-api}, all LLMs achieved better results than the pre-existing tests.

\subsection{RQ2: \RQTwo}\label{subsec:resultsRQ2}

To determine the most efficient LLMs, we computed a final score (average calculated score) for each model, based on a weighted combination of success rate, coverage, and mutation score using the TOPSIS technique. The results, ordered from highest to lowest, are presented in Table~\ref{tab:results}.

The top-performing models were \allBestLLMs[mv], with calculated scores ranging from \llm{s4}[S]\% to \llm{s1}[S]\%. Among them, \llm{s1}[mv] stood out as the most efficient, ranking first in all metrics and being the only model that produced no failed tests (Table~\ref{tab:fails}). We analyze however that the models \otherBestLLM[mv] achieved results close to \bestLLM[m] in all metrics, with a maximum difference of \diffMaxAllMetricsBetweenFirstAndOtherBestLLMs\%. The ranking of models by individual metrics and their relative differences from the top performer is shown in Table~\ref{tab:ranking-by-metric}. 

\newcommand{\rankingTableContent}{} 
\newcommand{\rankingTableLine}[1]{%
  {#1}° & \llm{s#1}[m] & \llm{s#1}[DS]\%     
        & \llm{p#1}[m] & \llm{p#1}[DP]\%     
        & \llm{c#1}[m] & \llm{c#1}[DC]\%      
        & \llm{m#1}[m] & \llm{m#1}[DM]\%  \\
}

\makeatletter
\foreach \index in \llmsIndexEnum {
  \edef\temp{\noexpand\g@addto@macro\noexpand\rankingTableContent{%
    \noexpand\rankingTableLine{\index}%
  }}
  \temp
}
\makeatother

\begin{table*}[ht]
  \centering
  \caption{Models performance ranking by metric, and difference ($\Delta$) from first position}
  \label{tab:ranking-by-metric}
  \begin{tabular*}{\textwidth}{@{\extracolsep{\fill}}llllllllll}
    \toprule
    Position & \( S \) & $\Delta$\( S \) 
             & \( SR \) & $\Delta$\( SR \) 
             & \( C \) & $\Delta$\( C \) 
             & \( M \) & $\Delta$\( M \) \\
    \midrule
    \rankingTableContent
    \bottomrule
  \end{tabular*}
\end{table*}

Although the LLMs \worstLLMs[mv] presented lower average scores, they still achieved solid results, particularly in success rate and, in some cases, coverage or mutation score. For example, \llm{gemini}[m] ranked second in coverage with only \llm{gemini}[DC]\% below the best model. Similarly, \llm{qwen}[m] and \llm{gpt}[m] showed competitive mutation scores with differences of only \llm{qwen}[DM]\% and \llm{gpt}[DM]\% respectively. Nevertheless, all models demonstrated satisfactory performance overall, with average calculated score values ranging between \minFinalScore\%, and \maxFinalScore\%, a difference of at most \diffMinMaxFinalScore\%.

In addition to performance, we also tracked the average total cost of processing each project with each LLM. As shown in Table~\ref{tab:results}, the execution cost per project remained very low across all models. Even the highest, \llm{deepseek}[m], remained under one dollar (\$\llm{deepseek}[TC]), while models like \llm{llama}[m], \llm{sabia}[m], and \llm{qwen}[m] stood out for delivering competitive results at extremely low costs—each below \$\llm{qwen}[TC] per execution. This reinforces the feasibility of adopting LLMs for test generation even in budget-constrained environments.

\subsection{Analysis of test failures}\label{subsec:analysisfailures}

Out of the \totalTests~ tests generated, \totalFailedTests~ failed to execute correctly, representing \totalFailedTestsRate\%. Table~\ref{tab:fails} details the number of failed tests per model. All models except \llm{claude}[mv] produced at least one faulty test. Upon manual inspection of the failures, we identified six recurring categories of errors, as shown below:

\begin{itemize}
    \item \textbf{15 failures – Property Length}: Validation of boundary values outside allowed ranges.
    \item \textbf{7 failures – Misinterpretation}: Incorrect logic or misunderstanding of the API specification.
    \item \textbf{5 failures – Authentication}: Missing or incorrect use of authentication.
    \item \textbf{5 failures – Property Requirement}: Misuse of required or optional fields.
    \item \textbf{4 failures – Required Characters}: Missing required characters (uppercase, digit, etc.).
    \item \textbf{3 failures – JSON Deserialization}: Errors in parsing response content.
\end{itemize}

These results reinforce the need for complementary techniques when using LLMs to improve test generation, to evolve from satisfactory results to exceptional results, whether in optimizing inputs, presented examples, prompt engineering or better models.

\newcommand{\failsTableContent}{} 
\newcommand{\failsTableLine}[1]{%
  \llm{#1}[mv] & \llm{#1}[TT] & \llm{#1}[FT] & \llm{#1}[PFT]\% \\
}

\makeatletter
\foreach \model in \llmsScoreEnum {
  \edef\temp{\noexpand\g@addto@macro\noexpand\failsTableContent{%
    \noexpand\failsTableLine{\model}%
  }}
  \temp
}
\makeatother

\begin{table}[ht]
  \caption{Failed tests produced by the LLMs}
  \label{tab:fails}
  \begin{tabular}{lccc}
    \toprule
      {\small Model } &
      {\small Total of Tests } & 
      {\small Failed Tests } & 
      {\small \% Failed Tests} \\
    \midrule
    \failsTableContent    
    \bottomrule
\end{tabular}
\end{table}


\newcommand{\ThreatsToValidityAndLimitations}{%
    T\MakeLowercase{hreats to validity and limitations}
}
\section{\ThreatsToValidityAndLimitations}\label{sec:limitations}

This experiment presents promising results in the use of LLMs for integration test generation, but some limitations, and threats to the validity of the study must be considered.

\textbf{LLM Selection:} The choice of LLMs was primarily based on model popularity. However, this may affect generalizability of our results, as emerging, lesser-known models, or those specifically designed for coding or trained for testing tasks, may yield different outcomes. We acknowledge this as a limitation and suggest that future studies include such models to broaden the analysis.

\textbf{Project Context:} The study focused on six open-source projects with relatively simple REST APIs, selected based on public availability and objective criteria. While suitable for evaluating the proposed method, more complex industrial systems—with larger  or asynchronous APIs, stricter security constraints, or broader business logic—may present challenges not covered here. We note this as a limitation in the applicability to enterprise-scale scenarios.

\textbf{OpenAPI Dependency:} Our approach depends on the availability and quality of OpenAPI specifications to drive test generation. Although OpenAPI is widely adopted, some legacy systems may lack such documentation or provide outdated or incomplete specifications, which could hinder the effectiveness of the approach. Future adaptations may consider alternative specification formats or test extraction from codebases directly.

\textbf{Prompt Engineering Dependency:} The quality of the results is strongly influenced by prompt design.  Variations in prompt wording, structure, and examples can influence the output quality, as can the language used—Portuguese, in our case. Although prompt design was carefully iterated and validated across models, we recognize that this process introduces variability and may require adjustments in other languages or technologies.

\textbf{Result Randomness:} The use of a temperature value of 1 during LLM execution introduces stochasticity, meaning that repeated executions with the same prompts may produce different outputs. This affects reproducibility and consistency. We partially mitigated this by using a fixed seed where supported and by repeating executions during validation. Nonetheless, some models exhibited variations across runs and experiments with these variations are important to expand the knowledge and impact of this parameter.

\textbf{Limited Metric Evaluation:} While success rate, branch coverage, and mutation score are well-established metrics in software testing, they do not fully capture other aspects such as readability, maintainability, runtime performance, or test execution cost. These complementary metrics and aspects could be explored in future work to broaden the quality analysis of the generated tests.

\textbf{Subjectivity in Qualitative Analysis:} The qualitative assessment of test cases—particularly for RQ1—was performed manually by the first author. Despite their 10+ years of experience in software engineering and REST API testing, the evaluation is inherently subjective and may vary if performed by other reviewers. To reduce bias, all results were thoroughly cross-checked with the corresponding OpenAPI specifications and expected patterns.

\textbf{Generalization of Results:} The results reflect the behavior of the evaluated models within the context of the selected projects. While diverse, these projects may not represent all types of REST APIs or organizational contexts. As such, the findings may not generalize to environments with significantly different technical stacks or operational constraints.

\textbf{Tokenization Limit:} Models with limited context windows occasionally produced incomplete outputs due to prompt size. To address this, we segmented the generation by endpoint group, which improved consistency. However, larger projects may still require new strategies to manage prompt length effectively.

\vspace{-0.5em}

\newcommand{\Conclusion}{%
    C\MakeLowercase{onclusion}
}
\section{\Conclusion}\label{sec:conclusion}

This approach - RestTSLLM - proposed the use of TSL and LLM to automate the generation of integration tests for REST APIs from its OpenAPI specification, and its experimental evaluation assessed its effectiveness. A comparative evaluation of tools was conducted, defining specific prompts, and clear criteria for metric analysis, including success rate, coverage, and mutation score. 

The experimental results demonstrated that all evaluated LLMs were capable of generating integration tests with satisfactory quality, based on both subjective and objective evaluation criteria. Notably, the models \allBestLLMs[mv] achieved the best overall performance, with \bestLLM[mv] standing out as the top-performing model. It achieved the highest average calculated score, ranked first in all individual metrics, and was the only model that produced no faulty tests. These results reinforce the feasibility of using LLMs to automate integration testing for REST APIs and highlight the potential of prompt-based strategies combined with intermediate TSL generation to guide LLMs more effectively.

In summary, this study contributes by proposing a reusable approach -- RestTSLLM -- for LLM-based test generation supported by prompt engineering techniques. The comparison of multiple LLMs provided valuable insights into their relative effectiveness, while the development of an automated multi-model execution script lays the groundwork for future, large-scale experimentation across diverse models and configurations.


Future work includes addressing the limitations discussed throughout this study, exploring ways to strengthen and broaden the applicability of the proposed approach. First, we plan to conduct additional experiments to validate the generalizability of the method. This includes testing with other emerging LLM tools, applying the approach to more complex architectures (such as event-driven and microservices systems), and using large-scale REST APIs with pre-existing test suites to enable quantitative comparisons. We also aim to evaluate its applicability with different technologies beyond .NET and xUnit.

Second, we intend to enhance the methodology by developing new features that improve automation and usability. This includes automatic handling and correction of errors through reprompting techniques, combining multiple generated tests to optimize coverage and fault detection, implementing IDE plugins to better integrate the approach into developers' workflows, and supporting test maintenance and evolution throughout the software lifecycle, enabling incremental updates, improving support for outdated or incomplete API specifications, and incorporating alternative input formats beyond OpenAPI.

In addition, we plan to investigate the use of additional languages in prompts and examples—beyond Portuguese and English—to evaluate multilingual performance. We will also compare the time and effort of manual test creation versus LLM-assisted generation and study ways to mitigate LLM token limitations in scenarios involving long prompts.

Finally, our approach not only aligns with key stages of the software testing lifecycle for REST APIs—particularly test case preparation and implementation—but also demonstrates its potential to enhance and scale traditional testing practices. These results reinforce the feasibility and relevance of adopting LLMs as a viable strategy for advancing test automation in real-world scenarios.

\section*{Artifact availability}\label{sec:artifact}

The authors declare that the research artifacts supporting the findings of this study are accessible at \url{https://doi.org/10.6084/m9.figshare.29525768.v4}. The artifacts can also be found on GitHub, accessible at \url{https://github.com/uffsoftwaretesting/RestTSLLM}.

\begin{acks}
This paper has been supported by
CNPq - \textit{National Council for Scientific and Technological Development} (grants 420025/2023-5 and 307088/2023-5), FAPERJ - \textit{Fundação Carlos Chagas Filho de Amparo à Pesquisa do Estado do Rio de Janeiro}, processes SEI-260003/002930/2024, SEI-260003/000614/2023,  and Coordenação de Aperfeiçoamento de Pessoal de Nível Superior – Brasil (CAPES) – Finance Code 001.
\end{acks}

\bibliographystyle{ACM-Reference-Format}
\bibliography{sample-base}

\setcopyright{none}
\settopmatter{printacmref=false} 
\renewcommand\footnotetextcopyrightpermission[1]{} 

\end{document}